\documentclass[11pt]{article}
\usepackage{epsfig}
\usepackage[usenames]{color}
\newtheorem{propos}{}[section]
\newcommand{\bprop}{\begin{propos}}
\newcommand{\eprop}{\end{propos}}

\newcounter{Roman}
\addtocounter{Roman}{1}

\newcommand{\beq}{\begin{equation}}
\newcommand{\eeq}{\end{equation}}
\newcommand{\bea}{\begin{eqnarray}}
\newcommand{\eea}{\end{eqnarray}}
\newcounter{saveeqn}

\input epsf
\begin{document}

\hfill \vbox{\hbox{}} 
\begin{center}{\Large\bf Reply to arXiv:0711.4930[hep-th] by Ito and 
Seiler}\\[2cm] 
{\bf E. T. Tomboulis\footnote{\sf e-mail: tombouli@physics.ucla.edu}
}\\
{\em Department of Physics, UCLA, Los Angeles, 
CA 90095-1547}
\end{center}
\vspace{1cm}

\begin{center}{\Large\bf Abstract}
\end{center} In a recent note (arXiv:0711.4930[hep-th]) Ito and Seiler 
claim that there is a 'missing link' in the derivation in 
arXiv:0707.2179[hep-th] by the present author; namely, that no proof 
of a certain inequality used there is 
given at weak coupling. Here it is pointed out that in fact 
no such missing link is present. 
The argument in 0707.2179 is, among other 
things, specifically constructed so that the inequality in question 
is invoked {\it only} at 
strong coupling, where it is easily proven. 
Underlying the mangling of the argument in  
0707.2179 by Ito and Seiler are their incorrect 
statements concerning the dependence of the  potential-moving 
decimation procedures used in 0707.2179 on space-time dimensionality and 
other decimation parameters.

\section{Introduction and conclusion}

In a recent paper by this author \cite{TT1} a derivation 
is given of the existence in $SU(2)$ gauge theory of a confining phase for all 
values of the coupling. In \cite{IS}
K.R. Ito and E. Seiler claim to have found 'missing links' in this derivation.

Missing links in a complete argument can be trivially created by simply 
omitting parts of the argument. This is all that Ito and Seiler do 
in their note. 
Specifically, they focus on the proof of 
inequality (5.15) in \cite{TT1}, restated as (3.5) in \cite{IS}. 
This inequality is easily proven in the strong coupling regime, i..e 
the region of convergence of the strong coupling expansion. However, 
there is no proof of it in the weak coupling regime. This is their 
'missing link'. 

In fact, no such gap in the derivation is present: the inequality in 
question is invoked {\it only} at strong coupling. 
Indeed, the development leading 
up to the appearance of the inequality is 
laid out in explicit detail in the first part of section 5 (p. 24 - 26) 
in \cite{TT1}.  
It is specifically constructed so that, by first allowing for a sufficently 
large number of RG decimations, the RG flow enters the regime 
of the convergence of the strong coupling expansion. It is only at this 
point that the further development of the argument leads to 
invoking inequality (5.15). In other words, by  construction,  
inequality (5.15) does not come up at weak coupling (large $\beta$).

Though all this should be clear to anyone with even a cursory familiarity 
with the argument in \cite{TT1}, we go through the 
obligatory chore of spelling this out in some more detail in section 2 
below. The 
reader familiar with 
the argument in \cite{TT1} can skip this and proceed to the next  
section.    

Underlying the misrepresentation of the argument in \cite{TT1} by 
Ito and Seiler are various incorrect assertions in \cite{IS} 
concerning the workings of the potential-moving decimations employed 
in \cite{TT1}. This is discussed in section 3.

\section{The procedure followed in \cite{TT1}}
The presentation of the various derivation steps, as well as their motivation, 
is quite explicit in \cite{TT1}, 
and the interested reader is referred to it for full details. A brief outline 
of the basic steps is given in \cite{TT2}. Still, in view of the mangling by 
Ito and Seiler of 
the actual procedure laid out in \cite{TT1}, it will be helpful 
to briefly reiterate some relevant parts of it here. 

Starting from a given lattice action $A_p$, such as the Wilson action, on 
a lattice $\Lambda$, we work in terms of the character 
expansion  
$
\exp{A_p(U)} = F_0\,\Big[\, 1 + \sum_{j\not= 0} d_j\, c_j(\beta)\,
\chi_j(U)\, \Big] $. 
The partition function on $\Lambda$ is then defined by 
\beq
Z_\Lambda(\beta) = \int dU_\Lambda\;\prod_p\, 
\Big[\, 1 + \sum_{j\not= 0} d_j\, c_j(\beta)\,
\chi_j(U)\,  \Big] \equiv Z_\Lambda\Big(\{c_j(\beta)\}\Big)  \,.\label{PF1}
\eeq
Decimations are next introduced whereby the lattice spacing $a$ is 
changed by a scale factor $b$ in successive steps:  
$a \to b a \to b^2 a \to \cdots \to b^n a$, generating 
the corresponding decimated lattices: 
$\Lambda \to \Lambda^{(1)} \to \Lambda^{(2)} \to \cdots \to 
\Lambda^{(n)}$. 
Decimation operations of the 'potential-moving type' are employed,  
which preserve the form (\ref{PF1}). 
Such a decimation operation can be 
summarized as a set of   
rules for the computation of the  
coefficients of the character expansion at the $(m+1)$-th step in terms
of those of the $m$-th step: 
\bea 
F_0(m) & = & F_0(\zeta, b, \{ c_i(m-1)\}) \label{RG1} \\
c_j(m) & = & c_j(\zeta, r, b, \{ c_i(m-1)\}) \; . \label{RG2} 
\eea 
The explicit forms of (\ref{RG1}) - (\ref{RG2}) are given in  
\cite{TT1}, section 2. Here we 
only note that they involve parameters $\zeta$, $r$ which control the amount 
by which the plaquette interactions remaining after a decimation 
step are `renormalized' to compensate for the ones that were removed. 
Correspondingly, the partition function undergoes the transformation
\beq
Z_{\Lambda^{(m-1)}}\Big(\{c_j(m-1)\}\Big) \to 
F_0(m)^{|\Lambda^{(m)}|} \, Z_{\Lambda^{(m)}}\Big(\{c_j(m)\}\Big) \;.
\label{PF2}
\eeq 
There is a bulk free energy contribution from the blocking $b^{m-1} a \to
b^{m} a$, whereas $Z_{\Lambda^{(m)}}\Big(\{c_j(m)\}\Big)$ is the 
partition function, again of the the form (\ref{PF1}), on the 
resulting lattice $\Lambda^{(m)}$.

The basic idea underlying the development in \cite{TT1} 
is very simple. For the 
choice: $\zeta=b^{d-2}$, $\quad r=1-\epsilon\;, \;  
0\leq \epsilon < 1 \;$ of decimation parameters, 
the r.h.s. in (\ref{PF2}) gives an upper bound on 
the the l.h.s., i.e. on the partition function of the previous step. 
For other choices, such as  $\zeta=1$, $\quad r=1 \;$, the r.h.s. is 
a lower bound on the l.h.s.   
Introducing an interpolating  parameter 
$\alpha$, $0\leq \alpha\leq 1$, one then defines   
coefficients $\tilde{c}_j(m,\alpha,r)$ and 
$\tilde{F}_0(m,\alpha)$ interpolating between the upper and lower 
bound coefficients at $\alpha=1$ and $\alpha=0$, respectively.   
But since there is nothing unique about any one such interpolation, it is 
expedient to consider more generally a family of such smooth
interpolations parametrized 
by a parameter $t$ in some interval  $(t_1, t_2)$.  
It suffices here to introduce $t$ in $\tilde{F}_0(m,\alpha,t)$ 
(cf. \cite{TT1} for details).

Then the upper-lower bounds statement 
implies that, for each value of $t$, and any other parameters such as $r$,    
there exist some value of the interpolating parameter 
$0 < \alpha=\alpha^{(m)}_\Lambda(t,r) < 1$, 
such that  
\beq 
\tilde{F}_0(m,\alpha,t)^{|\Lambda^{(m)}|} \,  
Z_{\Lambda^{(m)}}\Big(\{\tilde{c}_j(m,\alpha,r)\}\Big) 
= Z_{\Lambda^{(m-1)}} \,.
\label{Ifix} 
\eeq
Note that, by construction, there is parametrization invariance under 
shift in t in the l.h.s. of (\ref{Ifix}); in other words,  
$\alpha^{(m)}_\Lambda(t,r)$ is the level surface fixed
by (\ref{Ifix}). This allows determination of the derivatives of  
$\alpha^{(m)}_\Lambda(t,r)$ w.r.t. $t$ and $r$. 
A considerable part of \cite{TT1} is devoted to 
various bounds and estimates concerning the behavior of these 
derivatives.

Iterating this procedure starting from the original lattice, one gets 
an exact integral representation of the partition function  
on successively decimated lattices: 
\beq 
Z_\Lambda(\beta) = 
 \Big[\prod_{m=1}^n 
\tilde{F}_0(m,\alpha_\Lambda^{(m)}(t_m),t_m)^{|\Lambda|/b^{dm}} \Big]
\,Z_{\Lambda^{(n)}}\Big(\{\tilde{c}_j(n,\alpha_\Lambda^{(n)}(t_n))\}\Big) 
\, .\label{A}
\eeq
To construct something that can serve as a long-distance order parameter 
one considers the partition function 
$Z_\Lambda^{(-)}\;$ in the presence of a vortex, which is 
introduced by shifting the action on each element 
of a coclosed plaquette set ${\cal V}$ by 
a non-trivial  element ($\tau=-1$) of the group center. 
To have reflection positivity in all planes in the presence of the 
flux one may simply 
replace $Z_\Lambda^{(-)}$ by $Z^+_\Lambda \equiv 
{1\over 2}\Big(Z_\Lambda + Z_\Lambda^{(-)}
\Big)$. 
The above development can then be carried through also for $Z^+_\Lambda$ 
applying the same decimations (\ref{RG1}) - (\ref{RG2}).   
One thus obtains the corresponding integral 
representation on successively decimated lattices:  
\bea 
Z^+_\Lambda 
& = & \Big[\prod_{m=1}^n 
\tilde{F}_0(m,\alpha_\Lambda^{+(m)}(t_m),t_m)^{|\Lambda|/b^{dm}} \Big] 
\nonumber\\
& &  \cdot\; {1\over 2} 
\Big[\,Z_{\Lambda^{(n)}}\Big(\{\tilde{c}_j(n,\alpha_\Lambda^{(+)}(t_n))\}\Big) 
+ Z_{\Lambda^{(n)}}^{(-)}\Big(\{\tilde{c}_j(n,\alpha_\Lambda^{+(n)}(t_n))\}
\Big) \Big]  \,.\label{B}
\eea
As can be seen from (\ref{B}), the external flux presence does not affect 
the bulk free-energy contributions that resulted from successive blockings. 
Also, as indicated by the notation,   
the values $\alpha_\Lambda^{+(m)}(t)$ fixed at each  step 
in this representation are a priori 
distinct  from the values 
$\alpha_\Lambda^{(m)}(t)$ in the representation (\ref{A}) for  
$Z_\Lambda(\beta)$.

The vortex free energy order parameter is defined as the ratio of 
${Z_\Lambda^{(-)}/Z_\Lambda}$ or equivalently 
${Z_\Lambda^+ / Z_\Lambda}$.   
One may now represent this ratio on successively decimated lattices 
by inserting our representations (\ref{A}), (\ref{B}) 
in the numerator and denominator in ${Z_\Lambda^+ / Z_\Lambda}$. 
To account for any small 
discrepancies between $\alpha_\Lambda^{+(m)}(t)$ and 
$\alpha_\Lambda^{(m)}(t)$, and arrive at the statement V.1 in \cite{TT1}, 
the following procedure is followed. 

One first utilizes the independent invariance under $t$-parametrization 
shifts in numerator and denominator to arrange for cancellation 
of the bulk free energy pieces due to the $\tilde{F}_0$-coefficients 
(trivial characters) between numerator and denominator 
generated at each successive decimation step.  Thus after one decimation: 
\bea 
{Z_\Lambda + Z_\Lambda^{(-)}\over Z_\Lambda} 
& = & 
{\tilde{F}_0(1,\alpha_\Lambda^{+(1)}(t_1),t_1)^{|\Lambda^{(1)}|} 
\over \tilde{F}_0(1,\alpha_\Lambda^{(1)}(t_1),t_1)^{|\Lambda^{(1)}|} }
\nonumber\\
& &  \cdot\; {
\Big[\,Z_{\Lambda^{(1)}}\Big(\{\tilde{c}_j(1,\alpha_\Lambda^{(+)}(t_1))\}\Big) 
+ Z_{\Lambda^{(1)}}^{(-)}\Big(\{\tilde{c}_j(1,\alpha_\Lambda^{+(1)}(t_1))\}
\Big) \Big]  \over  
Z_{\Lambda^{(1)}}\Big(\{\tilde{c}_j(1,\alpha_\Lambda^{(+)}(t_1))\}\Big)  }
\nonumber \\
& = & 
{
\Big[\,Z_{\Lambda^{(1)}}\Big(\{\tilde{c}_j(1,\alpha_\Lambda^{(+)}(t_1))\}\Big) 
+ Z_{\Lambda^{(1)}}^{(-)}\Big(\{\tilde{c}_j(1,\alpha_\Lambda^{+(1)}(t_1))\}
\Big) \Big]  \over  
Z_{\Lambda^{(1)}}\Big(\{\tilde{c}_j(1,\alpha_\Lambda^{(+)}(t^{\;\prime}_1))\}
\Big)  }
\eea
by starting with a common $t_1=t_1^+$ in numerator and denominator, 
and shifting $t_1 \to 
t_1^{\;\prime}$ in the denominator so that 
$\tilde{F}_0(1,\alpha_\Lambda^{(1)}(t_1^\prime),t_1^\prime)^{|\Lambda^{(1)}|}
= \tilde{F}_0(1,\alpha_\Lambda^{+(1)}(t_1),t_1)^{|\Lambda^{(1)}|}$.  
This is always possible as long as the derivatives 
of $\alpha_\Lambda^{(m)}(t,r)$ and $\alpha_\Lambda^{+(m)}(t,r)$ w.r.t. $t$ 
do not vanish, or more precisely are bounded from below by a non-zero 
constant independent of the lattice volume. 
As discussed in detail in \cite{TT1},  
this can always be ensured by taking the
decimation parameter $r< 1$. 

Carrying out  $n$ successive decimation steps in this manner, 
one ends up with
\bea 
{Z_\Lambda + Z_\Lambda^{(-)}\over Z_\Lambda} 
& = & 
{\tilde{F}_0(n,\alpha_\Lambda^{+(n)}(t^+_n),t^+_n)^{|\Lambda^{(n)}|} 
\over \tilde{F}_0(n,\alpha_\Lambda^{(n)}(t_n),t_n)^{|\Lambda^{(n)}|} }
\nonumber\\
& &  \cdot\; {
\Big[\,Z_{\Lambda^{(n)}}\Big(\{\tilde{c}_j(n,\alpha_\Lambda^{+(n)}(t^+_n))\}
\Big) 
+ Z_{\Lambda^{(n)}}^{(-)}\Big(\{\tilde{c}_j(n,\alpha_\Lambda^{+(n)}(t^+_n))\}
\Big) \Big]  \over  
Z_{\Lambda^{(n)}}\Big(\{\tilde{c}_j(n,\alpha_\Lambda^{(n)}(t_n))\}\Big)  }
\,.\label{R2}
\eea
Take $n$ such that the flow under the sucessive decimations 
has entered the strong coupling regime, i.e. let $n$ be 
{\it sufficiently large}.   
At this point one is ready to cancel not only the bulk contributions 
from the $\tilde{F}_0$ coefficients but also those in the second 
factor in (\ref{R2}), i.e. from all scales. This can be done if 
one can find a value $t_n=t^+_n=t^*_\Lambda$ such that 
$\alpha_\Lambda^{(n)}(t^*)=\alpha_\Lambda^{+(n)}(t^*)
\equiv \; \alpha_\Lambda^{*(n)}$. As shown in \cite{TT1}, and also summarised  
in \cite{IS}, such a $t^*_\Lambda$ exists provided the inequality 
(5.15) in \cite{TT1} ( restated as (3.5) in \cite{IS}) holds. 
This is a comparison inequality between the derivative w.r.t. 
the interpolation parameter $\alpha$ of the log of the partition function 
with versus that without the vortex flux. 
It is an easy matter to 
demonstrate that indeed it holds within the strong coupling regime. 
The result V.1 in \cite{TT1} (which is incorrectly restated as 
`alleged theorem' or 
claim 2.1 with  $n=1$ in \cite{IS}, p. 5) 
then follows, as well as all the consequences spelled out in detail 
in \cite{TT1}.

\section{The Ito-Seiler miss of the non-missing link} 
As we just saw inequality (5.15) in \cite{TT1} (restated as 
(3.5) in \cite{IS}) is invoked only after the decimation 
flow has run into the region of convergence of the strong coupling expansion. 

Ito and Seiler make the following curious statement: 

``$\ldots$  inequality (3.5), which is, as far as we can see, not proven, 
even though 
the author remarks at the beginning of p. 28 of \cite{TT1}: 

``Assume now that under successive decimations the coefficients
$c^U_j (m)$ evolve within the convergence radius $\ldots$. Taking then
$n$ sufficiently large, we need establish inequality (5.15) (namely
$A \geq A^+$) only at strong coupling.'' `` 

These statements in \cite{TT1} quoted by them 
were, of course, made in the context of 
laying out the procedure, outlined in the last part of 
the previous section, that 
results in the inequality being invoked 
only at strong coupling. 
But, despite the explicit quotations, 
Ito and Seiler proceed to completely ignore the development that led to them. 
Instead, they simply state (their italics): 

``{\it It is not clear where and how his claim}, (i.e. the inequality 
in question), 
{\it is proven for large $\beta$ where the high-temperature expansion 
never works!} `` 
It is worth noting here that they never indicate the point in \cite{TT1} 
at which the inequality is needed at large $\beta$. There is, 
of course, no such point.

To reiterate, {\it there was never any issue of proving 
the inequality at large $\beta$; the argument in \cite{TT1} 
was devised so that 
the inequality need not be invoked at large $\beta$}. It is, in fact, not  
a priori clear whether 
it holds at large $\beta$.  At strong coupling, on the other hand, 
where it {\it is} invoked, the  proof is immediate, 
as Ito and Seiler also state.

This misrepresentation of the argument in \cite{TT1} by Ito and Seiler 
appears to originate in their incorrect assertions concerning  the 
RG decimations of the potential-moving type that are employed in \cite{TT1}. 
On p. 6 (Remark 2.2 (2)) they state: 

``The parameter $r$ increases the dimension $D$ 
from $D = 4$ to $D\geq 4$ from the 
point of view of the renormalization groups. So we set $r = 1$ in this paper. 
The introduction of r does not change our argument.'' 
 
From the point 
of view of which 'renormalization groups'?
We are talking here about precisely specified decimation prescriptions 
(section 2 in \cite{TT1}) which are then used to derive various statements. 
In terms of these, the quoted statement is manifestly 
incorrect, and in fact absurd. It leads to their outright distortion of 
the chain of argument in \cite{TT1}. 

In implementing the plaquette moving decimation operations  
that allow various bounds to be derived, and 
maintaining  reflection positivity (positivity of the character expansion 
coefficients), 
spacetime  dimensionality enters through the actual spacetime lattice 
on which these operations are carried out, and can only be integer, and 
through the integer-valued parameter $\zeta$ (cf \cite{TT1}).  
The statement that changes in the spacetime dimensionality 
can be absorbed in (small)  
$r$-shifts, which renormalize 2-dimensional integrations in 
hypercell boundaries after the plaquette-moving operations, 
is non-sensical within the framework of the decimation operations 
defined in \cite{TT1}.\footnote{It is 
incorrect also outside this 
framework, i.e.  
if one considers non-integer dimensions by continuing the decimation 
transformation formulas -- which cannot be interpreted in terms of 
plaquette moving operations that form the basis of the bounds
in \cite{TT1}. Then  both  positive and negative expansion 
coefficients occur, and letting $r$ different from unity 
certainly cannot be said to amount to 
raising the spacetime dimension even in a vague qualitative sense, 
let alone as an algebraically correct statement.}   

$r$ is in fact another decimation parameter, one among several 
that can be introduced in the class of decimations of 
`the potential-moving' type, which 
includes the Migdal-Kadanoff ones as a special case. 
Allowing variation of this parameter is an integral part of the argument 
in \cite{TT1} since such variation ensures that 
derivatives of the level-surfaces of
$\alpha_\Lambda^{(n)}(t,r)$, $\alpha^{+(n)}_\Lambda(t,r)$ do not vanish as 
mentioned above. Without this assurance, the argument cannot go through. 
As pointed out in \cite{TT1}, another important outcome of this is to 
determine on which side of the MK choice of parameters the 
upper bound parameter values in \cite{TT1} must lie.

Despite casting everything in the format of definitions, theorems etc, 
in the usual pseudo-mathematical pretense at `rigor', 
general imprecision and confusion characterizes \cite{IS}  as 
manifested by the above quoted statements.\footnote{Various other odd 
comments appearing in \cite{IS}, such as the one on positivity 
of the character expansion coefficients and reflection positivity 
(p.5, remark 2.1 (1)),  
are of no overall importance to warrant comment.} 
The authors do not 
argue from detailed consideration of the complete argument 
in \cite{TT1}, parts of which they simply ignore; or from any 
knowledge gained by the actual computation of the decimation 
flows for a range of decimation parameters, spacetime 
dimensionalities and gauge 
groups. They in effect argue by a priori conviction, as well  
exemplified by their remarks following what they label claim 2.1 (p. 5-6),  
their misguided discussion section (p. 9), 
and their reminding 
us (their reference [2]) that at least one of them has long 
been advocating the absence of asymptotic 
freedom and its consequences in non-abelian gauge theory.

\end{document}